\documentstyle[aps,prl,twocolumn,floats,graphicx]{revtex}

\begin{document}
\draft
\preprint{}
\wideabs{
\title{Measurement of Transport Lifetime in a Cuprate Superconductor}
\author{J. Corson, J. Orenstein}
\address{Materials Sciences Division, Lawrence Berkeley National Laboratory and Physics Department, University
of California, Berkeley, California 94720}
\author{Seongshik Oh, J. O'Donnell, J. N. Eckstein}
\address{Department of Physics, University of Illinois, Urbana, Illinois 61801}
\date{\today}
\maketitle
\begin{abstract} 
We report measurements of the phase of the conductivity, $\phi_\sigma\equiv arg(\sigma)$, in the normal state of a $Bi_{2}Sr_{2}CaCu_{2}O_{8+\delta}$ (BSCCO) thin film from 0.2-1.0 THz.  From $\phi_\sigma$ we obtain the time delay of the current response, $\tau_\sigma\equiv\phi_\sigma/\omega$.  After discovering a systematic error in the data analysis, the extracted $\tau_\sigma$ has changed from that reported earlier.  The revised data is shown in the sole figure below.  Analysis and discussion of these data will follow.  
\end{abstract}
}

\begin{figure}[h]
     \includegraphics[width=3.25in]{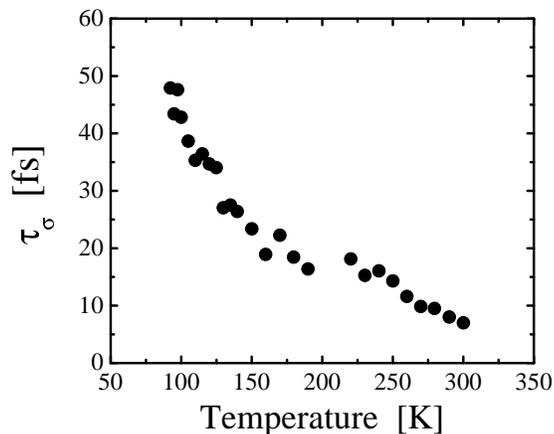}
\caption{$\tau_\sigma$ plotted versus temperature as black circles.  The error bars are the standard deviation in $\tau_\sigma$ obtained from 40 repeated one minute scans taken at room temperature.}
\label{fig:Third}
\end{figure}
This work was supported under NSF Grant No. 9870258, DOE Contract No. DE-AC03-76SF00098, and ONR Contract No. N00014-94-C-001. 

\end{document}